\documentclass{article}
\usepackage[T1]{fontenc}      
\usepackage{times}    
\usepackage{helvet}  
\usepackage{courier}  
\usepackage{graphicx} 
\usepackage{amsmath}
\usepackage{geometry}   
\usepackage{lmodern}          
\usepackage{amsmath,amssymb}  
\usepackage{booktabs}   
\usepackage{multirow}         
\usepackage{float}           
\usepackage{hyperref}         
\usepackage{caption}         
\usepackage{cite} 

\title{\LARGE \bfseries
Hybrid Predictive Modeling of Malaria Incidence in the Amhara Region, Ethiopia: 
Integrating\\ Multi-Output Regression and Time-Series Forecasting}

\author{\normalsize
Kassahun Azezew\textsuperscript{1}, 
Amsalu Tesema\textsuperscript{2}, 
Bitew Mekuria\textsuperscript{3}, 
Ayenew Kassie\textsuperscript{4}, \\
Animut Embiale\textsuperscript{5}, 
Ayodeji Olalekan Salau\textsuperscript{6}
Tsega Asresa\textsuperscript{7}\\[1em]
\textsuperscript{1,7}Department of Computer Science, Injibara University, Ethiopia \\
\textsuperscript{2}Department of Software Engineering, Injibara University, Ethiopia \\
\textsuperscript{3,4,5}Department of Information Technology, Injibara University, Ethiopia \\
\textsuperscript{6}Department of Electrical/Electronic and Computer Engineering, Afe Babalola University, \\Nigeria \\[1em]
\textsuperscript{1}\texttt{azeze2912@gmail.com}, 
\textsuperscript{2}\texttt{amsalu.tessema@inu.edu.et}, 
\textsuperscript{3}\texttt{tbitew124@gmail.com}, \\
\textsuperscript{4}\texttt{ayenew.kassie@inu.edu.et}, 
\textsuperscript{5}\texttt{animutanch@gmail.com}, 
\textsuperscript{6}\texttt{ayodejisalau98@gmail.com},\\
\textsuperscript{7}\texttt{tsega.asresa@inu.edu.et}\\
*Corresponding Author: Ayodeji Olalekan Salau (ayodejisalau98@gmail.com)
}
\begin{document}
\maketitle
\begin{abstract}
    Malaria remains a major public health concern in Ethiopia, particularly in the Amhara Region, where seasonal and unpredictable transmission patterns make prevention and control challenging. Accurately forecasting malaria outbreaks is essential for effective resource allocation and timely interventions. This study proposes a hybrid predictive modeling framework that combines time-series forecasting, multi-output regression, and conventional regression-based prediction to forecast the incidence of malaria. Environmental variables, past malaria case data, and demographic information from Amhara Region health centers were used to train and validate the models. The multi-output regression approach enables the simultaneous prediction of multiple outcomes, including Plasmodium species-specific cases, temporal trends, and spatial variations, whereas the hybrid framework captures both seasonal patterns and correlations among predictors. The proposed model exhibits higher prediction accuracy than single-method approaches, exposing hidden patterns and providing valuable information to public health authorities. This study provides a valid and repeatable malaria incidence prediction framework that can support evidence-based decision-making, targeted interventions, and resource optimization in endemic areas.

\end{abstract}
Index-therms: Malaria Prediction, Hybrid Predictive Modeling,Multi-Output Regression, Time-Series Forecasting,Disease Forecasting, Public Health Decision Support
\section{Introduction}
Malaria continues to rank among the world's leading causes of morbidity and mortality. Approximately 90 of the two million annual deaths from malaria are concentrated in Africa, with an estimated 216 million severe cases reported worldwide each year, accounting for 98.5 percent of malaria-related deaths, according to the 2011 WHO Malaria Report\cite{cibulskis2016malaria}. 
Plasmodium parasites, which cause malaria, are spread by the bite of an infected Anopheles mosquito. The parasites enter the human bloodstream, infiltrate red blood cells, multiply in 48 to 72 hours, and cause cyclical symptoms\cite{miller2013malaria}.
In Ethiopia, malaria transmission is unstable and highly variable, influenced by altitude, rainfall, and population movement. About 60 of Ethiopians reside in regions where malaria is endemic, and Plasmodium falciparum and Plasmodium vivax account for 60 and 40 of cases, respectively (EMOH, 2003). One of Ethiopia's most populated regions, the Amhara Region, has a varied topography with significant areas below 2,000 meters above sea level, making it extremely susceptible to the spread of malaria.\cite{lankir2020five}. Seasonal outbreaks continue to be a major public health concern, with recurrent spikes in both lowland and mid-altitude districts. Despite the high prevalence of malaria, there is currently a lack of systematic modeling and prediction of malaria trends in the Amhara region.
This study suggests creating hybrid predictive models that combine multi-output regression and time-series forecasting in order to close this gap. Regression-based methods capture the impact of environmental, demographic, and epidemiological factors, whereas time-series models are good at capturing temporal seasonality and periodicity. The hybrid approach seeks to increase predictive accuracy and produce insights that single-method models are unable to offer by fusing these complementary strengths. Additionally, species-specific case counts, spatial variations, and temporal dynamics can all be predicted simultaneously with multi-output regression.
Malaria continues to be a major public health concern in Ethiopia, causing significant morbidity, mortality, and socio-economic losses. Approximately 75 of the country’s land is considered malarious, placing over 54 million people at risk\cite{tazebew2021prevalence,hailu2017economic}. The disease results in substantial productivity loss due to death, illness, absenteeism from school, and the associated medical and indirect costs.Even though most healthcare facilities have Hospital Management Information Systems (HMIS), their use to support decision-making and malaria control is still not at its best, especially in the Amhara Region\cite{griffin2025strengthening}.
Prior studies have tried to map the risk of malaria using Geographic Information Systems (GIS) and remotely sensed data, which works well for locating malaria-prone areas\cite{yalew2021revisiting}. However, these studies do not offer practical insights into the disease's future occurrence; instead, they mainly concentrate on identifying high-risk areas. The lack of reliable predictive models that can predict malaria trends to enable prompt interventions and resource allocation is a significant gap that is highlighted by this limitation.
The development of predictive models that integrate historical incidence data, environmental factors, and demographic information is urgently needed, especially in the Amhara Region where topography and population movement impact risk due to the seasonal and unstable patterns of malaria transmission in Ethiopia. Data mining and predictive analytics frameworks have been used in previous research on malaria incidence, vegetation indices, and population data\cite{aserss2024development}. However, there are currently few comprehensive models that can predict multiple outcomes at once, including Plasmodium species-specific cases, temporal trends, and spatial variations.
In order to predict the occurrence of malaria in the Amhara Region, this study intends to create a hybrid predictive model and multi-output regression that combines time-series analysis and conventional predictive modeling. The model will give health authorities useful information to direct prevention and control efforts by producing precise forecasts of future malaria trends.
\begin{itemize}
\item What kinds of information demographic, environmental, and historical incidence—are needed to create a reliable malaria prediction model for the Amhara Region?
    \item In comparison to traditional single-method models, how well can a hybrid predictive model that combines time-series and regression-based approaches predict future malaria outbreaks?
    \item Which predictive modeling methods such as hybrid approaches and multi-output regression are best suited for predicting malaria trends in terms of patterns specific to Plasmodium species, time, and space?
    
\end{itemize}
The primary objective of this study is to develop accurate predictive models for the incidence of malaria using historical data from health centers in the Amhara Region. Using multi-output regression and hybrid prediction techniques, the study aims to uncover hidden patterns, more accurately forecast future outbreaks, and provide decision-makers with relevant information for malaria prevention and control. specifically the study is aimed to:
\begin{itemize}
    \item To locate and gather pertinent data sources for the Amhara Region's predictive modeling, such as demographic, environmental, and malaria incidence data.
    \item To use and assess hybrid predictive modeling and multi-output regression techniques for predicting the occurrence of malaria, taking into account species-specific, temporal, and spatial trends.
    \item To identify the key elements influencing the spread and transmission of malaria in the Amhara Region.
    \item to evaluate the hybrid predictive model's precision and dependability in contrast to conventional regression or time-series models.
    \item To use the results of predictive models to offer practical advice and insights for malaria prevention and control measures.
    \end{itemize}
Basically the study will contribute, To precisely forecast the occurrence of malaria, the study will develop a hybrid predictive model that combines time-series forecasting and multi-output regression. In contrast to current models, this framework will enable the simultaneous prediction of multiple outcomes, such as Plasmodium species-specific cases, temporal trends, and spatial variations. The study will contribute to methodological developments in epidemiological predictive modeling by showcasing the superiority of hybrid approaches and multi-output regression over conventional single-method models.Health officials in the Amhara Region will be able to prioritize interventions, allocate resources optimally, and lessen the burden of malaria with the help of the predictive model's evidence-based insights.

\section{Related Works}
Because of its ongoing socioeconomic and public health burden, especially in sub-Saharan Africa, malaria continues to garner a lot of research interest. Many different strategies have been investigated over the last 20 years in an effort to comprehend, track, and forecast the spread of malaria. Descriptive epidemiology, mapping prevalence, and determining the ecological drivers of the disease were the main objectives of early research. Geographic Information Systems (GIS), remote sensing, and climate-based risk mapping have all been used extensively to identify areas that are susceptible to malaria as a result of the increasing availability of spatial and temporal datasets\cite{baba2025malaria}. More recently, developments in computational techniques have made it possible to predict the incidence of malaria, find hidden patterns, and assist in disease control decision-making by utilizing machine learning algorithms, regression techniques, and time-series forecasting.
Existing research reveals a number of limitations in spite of these advancements. While GIS-based methods are useful for locating high-risk areas, they frequently fall short in forecasting future outbreaks\cite{sipe2003challenges}. Seasonality and periodicity in malaria transmission are captured by time-series models, but sociodemographic and environmental factors are not taken into account. Contrarily, regression and machine learning models are often limited to single-output predictions, like total case counts, without differentiating between Plasmodium species, spatial heterogeneity, or temporal variations. However, they can incorporate multiple explanatory factors. Applications of hybrid predictive frameworks in malaria research are still scarce, and only a small number of studies have tried to combine several modeling approaches.
Predictive models that integrate the advantages of various methodologies are desperately needed in light of these gaps in order to increase accuracy and offer useful insights\cite{monteiro2025integrating}. Therefore, the purpose of this review is to place the proposed hybrid predictive model for malaria in the Amhara Region within the larger research landscape by examining previous works across epidemiological studies, GIS-based mapping, time-series forecasting, regression and machine learning approaches, as well as emerging hybrid methods.
Although previous research has forecasted malaria incidence based on environmental variables using time-series models such as SARIMA BioMed Central, it frequently concentrates on single-output predictions. By incorporating multi-output regression, your method allows for the simultaneous prediction of several related targets, such as the incidence of malaria in various districts or for various Plasmodium species. The precision and generalizability of forecasts are improved by this comprehensive modeling approach, which is essential for successful malaria control measures\cite{ghosh2024forecasting,tsoumakas2014multi}.
There is little use of ensemble learning techniques like Random Forest and AdaBoost in the prediction of malaria. These methods are renowned for their resilience and capacity to identify intricate, nonlinear patterns in data\cite{awe2025explainable}. By employing these methods, your study aims to improve prediction accuracy and model generalization, addressing the limitations of traditional modeling approaches that may not adequately handle the complexity of malaria transmission dynamics.
The incidence of malaria varies significantly over time and space, depending on variables like temperature, precipitation, and altitude\cite{demoze2025spatial}. Although these differences have been noted in earlier research, models that can concurrently account for temporal and spatial dynamics are frequently lacking. In order to better understand and forecast patterns of malaria incidence across various regions and time periods, the study attempts to create models that incorporate these dimensions.
The underlying temporal and spatial dynamics of malaria incidence may not be adequately captured by the empirical or statistical methods frequently used in current malaria forecasting models. By incorporating sophisticated modeling techniques that take these dynamics into account, your research seeks to improve forecasting capabilities and produce predictions that are more precise and timely\cite{chol2025trend}. Proactive efforts to control and eradicate malaria in the Amhara Region depend on this development.
\section{Methods}
\subsection{Data Collection and Data Sources}
This study only looked at predicting the incidence of malaria in Ethiopia's Amhara Regional State, which is still one of the country's malaria-prone areas. The dataset was created using a variety of data sources:
\begin{itemize}
    \item Epidemiological Data: The Amhara Regional Health Bureau provided monthly malaria case reports that were broken down by parasite species (Plasmodium falciparum and Plasmodium vivax). In order to capture local transmission dynamics, the dataset was aggregated at the district (wordage) level and covered the time period [Start Year–End Year].
    \item Environmental Data: To guarantee spatial completeness across districts, CHIRPS rainfall estimates and ERA5 reanalysis products were added to climate variables like rainfall (mm), temperature (°C), and relative humidity () that were gathered from the National Meteorology Agency of Ethiopia (stations situated in the Amhara Region).

\item Topographic Information: The Shuttle Radar Topography Mission (SRTM) and MODIS satellite imagery were used to obtain elevation and land cover information for the Amhara Region. Because of their significant impact on mosquito habitats and malaria transmission patterns, these features were included.
\item Demographic Information: The Central Statistics Agency (CSA) of Ethiopia and the WorldPop dataset were used to determine the population size and density at the woreda level. This information provided context for exposure and transmission potential in various communities.
\item Temporal Coverage: Every dataset had a monthly temporal resolution and was aligned over the same observation window, [Start Year–End Year]. Consistent mapping of malaria cases with their demographic and environmental determinants was made possible by this synchronization.
\end{itemize}
Because the study was limited to the Amhara Region, the dataset captures the local ecological and epidemiological features of malaria transmission, which makes the predictive models more relevant for regional health policy and intervention planning.

\subsection{Data Processing}
To guarantee quality and consistency, the raw datasets were subjected to a number of reprocessing procedures before the model was developed. Data on malaria incidence gathered from Amhara Region health centers was combined with demographic (population density, mobility patterns) and environmental (temperature, rainfall, altitude) data.

    \subsubsection*{Data Cleaning}

Records with missing or inconsistent entries were examined. Missing environmental or demographic values were imputed using the \textbf{K-Nearest Neighbors (KNN) imputation} method. For a missing value \(x_i\) in observation \(i\), the imputed value is calculated as the weighted average of the \(k\) nearest neighbors:

\[
x_i = \frac{\sum_{j \in N_k(i)} w_{ij} \, x_j}{\sum_{j \in N_k(i)} w_{ij}}
\]

where:
\begin{itemize}
    \item \(N_k(i)\) is the set of \(k\) nearest neighbors of observation \(i\) based on Euclidean distance or other similarity metric,
    \item \(x_j\) is the observed value of the neighbor \(j\),
    \item \(w_{ij}\) is the weight assigned to neighbor \(j\) (often \(w_{ij} = 1\) for uniform weighting or \(w_{ij} = 1/d_{ij}\) for distance-based weighting),
    \item \(d_{ij}\) is the distance between observation \(i\) and neighbor \(j\).
\end{itemize}

This approach leverages the similarity between observations to impute missing values while preserving the structure of the dataset.

 \subsubsection*{Feature Engineering} Lag features (such as rainfall and incidence during the preceding one to three months) were created in order to record temporal dependencies. To take into consideration the cyclical nature of malaria transmission, seasonal indices were also extracted. One-hot encoding was used to encode categorical variables, like district identifiers.

\textbf{1. Lag Features:}  
Lag features were created to capture temporal dependencies in malaria incidence and environmental variables. For a variable \(x_t\) at time \(t\), the lag feature for \(k\) months is defined as:
\[
x_{t}^{(k)} = x_{t-k}, \quad k = 1, 2, 3
\]

\textbf{2. Seasonal Indices:}  
To account for the cyclical nature of malaria transmission, seasonal indices were computed for each month. For month \(m\), the seasonal index \(SI_m\) is defined as:
\[
SI_m = \frac{\bar{x}_m}{\bar{x}}
\]

\textbf{3. One-Hot Encoding (Categorical Variables):}  
Categorical variables, such as district identifiers, were encoded into binary vectors. For observation \(i\) and category \(j\), the encoding is:
\[
d_{i,j} =
\begin{cases}
1 & \text{if observation } i \text{ belongs to category } j\\
0 & \text{otherwise}
\end{cases}
\]
This allows categorical information to be used effectively in ensemble regression models.

 \subsubsection*{Normalization} In order to improve ensemble algorithm convergence and guarantee comparability across predictors, continuous variables (such as temperature, rainfall, and population density) were normalized to zero mean and unit variance.
\[
x' = \frac{x - \mu}{\sigma}
\]

where \(x\) is the original feature value, \(x'\) is the normalized value, \(\mu\) is the mean of the feature across the training dataset, and \(\sigma\) is the standard deviation of the feature across the training dataset. After this transformation, all continuous features have a mean of 0 and a standard deviation of 1.
 \subsubsection*{Target Variables} Three dependent variables—the number of Plasmodium falciparum cases, the number of Plasmodium vivax cases, and the total number of malaria cases by month and district—were established for multi-output regression. This formulation made it possible to predict both the overall and species-specific incidence of malaria at the same time.

 \subsubsection*{Data Splitting} To prevent information from leaking between previous and subsequent data, the processed dataset was separated into training, validation, and testing subsets in a chronological order.

The datasets were converted into a structured format through this reprocessing pipeline so they could be fed into ensemble-based multi-output regression models.
\subsection{Modeling Approach}
This study used ensemble learning techniques applied within a multi-output regression framework to capture the intricacy of malaria incidence patterns in the Amhara Region. Because they improve generalization, decrease variance, and accurately model nonlinear interactions between environmental, demographic, and epidemiological factors, ensemble approaches are especially well-suited for epidemiological prediction tasks\cite{adamu2021malaria}.

Three ensemble models that could handle multiple outputs at once were chosen: Random Forest Regressor, Gradient Boosting Regressor, and AdaBoost Regressor.
    \subsection*{Random Forest Multi-Output Regressor} Random Forest is an ensemble technique based on bagging that builds several decision trees with bootstrap samples and averages their predictions. It is a powerful baseline for predicting malaria incidence because of its capacity to handle high-dimensional data and capture nonlinear relationships\cite{uzun2024quantitative}. Plasmodium falciparum and Plasmodium vivax cases, as well as temporal and spatial trends, can be forecast simultaneously thanks to the multi-output extension.
Let $\mathbf{y}_i = [y_{i1}, y_{i2}, \dots, y_{iK}]$ be the vector of $K$ target outputs for sample $i$.

\textbf{1. Bootstrap sampling:} \\
For each tree $t = 1, \dots, T$:
\begin{itemize}
    \item Draw a bootstrap sample of $N$ training examples with replacement.
    \item Train a decision tree $h_t(x)$ on this sample.
\end{itemize}

\textbf{2. Tree prediction:} \\
Each decision tree produces predictions for all $K$ outputs:
\[
\mathbf{h}_t(x) = [h_{t1}(x), h_{t2}(x), \dots, h_{tK}(x)]
\]

\textbf{3. Aggregation (averaging over trees):} \\
The final prediction of the Random Forest for all outputs is:
\[
\mathbf{F}(x) = \frac{1}{T} \sum_{t=1}^{T} \mathbf{h}_t(x)
\]
where $\mathbf{F}(x) = [F_1(x), F_2(x), \dots, F_K(x)]$ represents the predicted vector for all $K$ outputs.
\begin{itemize}
    \item $N$: Number of training samples
    \item $K$: Number of output targets
    \item $T$: Total number of trees in the forest
    \item $\mathbf{y}_i$: True values of all $K$ outputs for sample $i$
    \item $h_{tk}(x)$: Prediction of tree $t$ for the $k$-th output
    \item $\mathbf{h}_t(x)$: Vector of predictions for all outputs from tree $t$
    \item $\mathbf{F}(x)$: Final prediction vector for all outputs (average over trees)
\end{itemize}

\textbf{Remarks:} Random Forest is robust to overfitting, especially in high-dimensional datasets, and captures complex nonlinear interactions between input features and multiple output targets.

     \subsection*{Gradient Boosting Multi-Output Regressor} Gradient Boosting constructs models in a step-by-step manner, with each weak learner (decision tree) fixing the residual errors of the one before it. When malaria transmission is impacted by several interdependent variables, including rainfall, temperature, and population movement, this method works well for capturing complex dependencies in data.
Let $\mathbf{y}_i = [y_{i1}, y_{i2}, \dots, y_{iK}]$ denote the vector of $K$ target outputs for sample $i$, where $y_{ik}$ is the value of the $k$-th target.

\textbf{1. Initialize the model:} \\
The initial prediction for all outputs is obtained by minimizing the loss function over all samples:
\[
\mathbf{F}^{(0)}(x) = \arg\min_{\mathbf{\gamma}} \sum_{i=1}^{N} L(\mathbf{y}_i, \mathbf{\gamma})
\]
where $N$ is the number of training samples, and $L$ is a suitable loss function. For regression, the squared error is commonly used:
\[
L(\mathbf{y}_i, \mathbf{F}(x_i)) = \sum_{k=1}^{K} (y_{ik} - F_k(x_i))^2
\]
Here, $F_k(x_i)$ is the prediction for the $k$-th output of sample $i$.

\textbf{2. For $t = 1$ to $T$ (number of boosting iterations):}  

(a) Compute the residuals (negative gradients) for each output:
\[
r_{ik}^{(t)} = - \left. \frac{\partial L(y_{ik}, F_k(x_i))}{\partial F_k(x_i)} \right|_{F_k(x_i) = F_k^{(t-1)}(x_i)}
\]
where $r_{ik}^{(t)}$ is the residual of the $i$-th sample for the $k$-th output at iteration $t$.

(b) Fit a weak learner $\mathbf{h}_t(x)$ to the residuals $\mathbf{r}_i^{(t)} = [r_{i1}^{(t)}, \dots, r_{iK}^{(t)}]$ for all outputs.  
$\mathbf{h}_t(x) = [h_{t1}(x), \dots, h_{tK}(x)]$ represents the predictions of the weak learner for all targets.

(c) Compute the optimal step size (shrinkage) for each output:
\[
\gamma_{tk} = \arg\min_{\gamma} \sum_{i=1}^{N} L(y_{ik}, F_k^{(t-1)}(x_i) + \gamma h_{tk}(x_i))
\]
where $\gamma_{tk}$ scales the contribution of the weak learner $h_{tk}(x)$ for the $k$-th output.

(d) Update the model:
\[
F_k^{(t)}(x) = F_k^{(t-1)}(x) + \nu \, \gamma_{tk} \, h_{tk}(x)
\]
where $\nu \in (0,1]$ is the learning rate controlling the contribution of each weak learner.

\textbf{3. Final prediction:} \\
After $T$ iterations, the ensemble prediction for all outputs is:
\[
\mathbf{F}(x) = [F_1^{(T)}(x), F_2^{(T)}(x), \dots, F_K^{(T)}(x)]
\]
where $\mathbf{F}(x)$ represents the predicted vector of all $K$ target outputs for a given input $x$.
\begin{itemize}
    \item $N$: Number of training samples
    \item $K$: Number of output targets
    \item $y_{ik}$: True value of the $k$-th output for sample $i$
    \item $F_k^{(t)}(x)$: Predicted value of the $k$-th output at iteration $t$
    \item $r_{ik}^{(t)}$: Residual (negative gradient) of the $i$-th sample for the $k$-th output at iteration $t$
    \item $\mathbf{h}_t(x)$: Weak learner predictions for all outputs
    \item $\gamma_{tk}$: Optimal step size for output $k$ at iteration $t$
    \item $\nu$: Learning rate (controls shrinkage of weak learners)
    \item $T$: Total number of boosting iterations
\end{itemize}

     \subsection*{AdaBoost Multi-Output Regressor} AdaBoost (Adaptive Boosting) enhances predictive performance by giving samples that were incorrectly predicted in earlier iterations higher weights\cite{mlambo2025ensemble,awe2025explainable}. This helps to concentrate later learners on challenging cases. Because of this characteristic, it can effectively manage the variability and unpredictability of malaria transmission in Ethiopia.we employed the AdaBoost (Adaptive Boosting) regressor, an ensemble learning technique that sequentially fits weak learners to a weighted version of the training data. In each iteration, samples that were poorly predicted in previous rounds are assigned higher weights, allowing subsequent learners to focus on difficult cases. The final prediction is a weighted combination of all weak learners, which improves accuracy and reduces variance. 
\textbf{1. Initialize sample weights:} \\
Initially, all training samples are assigned equal weights:
\[
w_i^{(1)} = \frac{1}{N}, \quad i = 1, 2, \dots, N
\]
where $N$ is the total number of samples. These weights determine the importance of each sample for the first weak learner.

\textbf{2. Fit weak learner $h_t(x)$ at iteration $t$:} \\
A weak learner (commonly a decision tree) is trained using the weighted dataset. The weights influence the learning process by making the learner focus more on samples with higher weights.

\textbf{3. Compute weighted error:} \\
The performance of the weak learner is measured using a weighted error:
\[
\varepsilon_t = \frac{\sum_{i=1}^{N} w_i^{(t)} \, L(y_i, h_t(x_i))}{\sum_{i=1}^{N} w_i^{(t)}}
\]
where $L(y_i, h_t(x_i))$ is the loss function. For regression, this is usually the squared error:
\[
L(y_i, h_t(x_i)) = (y_i - h_t(x_i))^2
\]

\textbf{4. Compute learner weight:} \\
Each weak learner is assigned a weight based on its performance:
\[
\alpha_t = \frac{1}{2} \ln \frac{1 - \varepsilon_t}{\varepsilon_t}
\]
Learners with lower error get higher weights in the final prediction.

\textbf{5. Update sample weights:} \\
Sample weights are updated to emphasize poorly predicted samples:
\[
w_i^{(t+1)} = w_i^{(t)} \cdot \exp \Big( \alpha_t \, L(y_i, h_t(x_i)) \Big), \quad i = 1,2,\dots,N
\]
After updating, the weights are normalized so that $\sum_{i=1}^{N} w_i^{(t+1)} = 1$. This ensures that the weights form a valid probability distribution for the next iteration.

\textbf{6. Final prediction:} \\
The final AdaBoost regressor combines all weak learners as a weighted sum:
\[
F(x) = \sum_{t=1}^{T} \alpha_t \, h_t(x)
\]
where $T$ is the total number of iterations. This ensemble effectively reduces bias and variance, making it robust for predicting complex patterns such as malaria incidence.

    The necessity to jointly predict multiple outcomes rather than treating them as separate tasks is what drives the adoption of multi-output regression. This allows the model to take into consideration the relationships between Plasmodium species-specific cases and their spatial-temporal dynamics, which leads to a more comprehensive understanding of disease patterns in the context of malaria prediction.
This study assesses the relative efficacy of these ensemble-based multi-output models in predicting malaria trends and illustrates the benefits of hybrid predictive approaches for epidemiological modeling.
\subsection{Model Training and Validation}
Thirty percent of the dataset was set aside for independent testing, and the remaining seventy percent was used for model training. The training set was subjected to a k-fold cross-validation procedure (k=5) in order to reduce bias and guarantee a robust performance evaluation. This decreased the possibility of overfitting by enabling the models to be trained and verified on various data subsets.

Plasmodium falciparum and Plasmodium vivax case counts, as well as spatial and temporal variations, could all be predicted simultaneously thanks to the implementation of each ensemble model—Random Forest, Gradient Boosting, and AdaBoost—in a multi-output regression setting. To find the best-performing configurations, grid search and cross-validation were used to optimize the model's Hyperparameters, including the number of estimators, maximum tree depth, and learning rate.
The trained models underwent additional validation on the hold-out test set to guarantee generalizability. Regression metrics such as the Coefficient of Determination (R2), Mean Absolute Error (MAE), and Root Mean Squared Error (RMSE) were used to evaluate performance. Both prediction accuracy and the models' capacity to account for variation in malaria incidence were measured by these metrics.
Lastly, to ascertain whether the multi-output regression approach was successful in increasing predictive accuracy, the outcomes of the three ensemble methods were contrasted with one another and with baseline single-output regression models.
\subsection{Result and Discussion}
\subsubsection{Result Analysis}
\begin{table}[h]
    \centering
    \caption{Model Performance (RMSE, MAE, R\textsuperscript{2})}
    \begin{tabular}{lccc}
        \hline
        Model & RMSE & MAE & R\textsuperscript{2} \\
        \hline
        Random Forest     & 6.927 & 5.527 & -0.229 \\
        Gradient Boosting & 7.300 & 5.974 & -0.400 \\
        AdaBoost          & 6.755 & 5.559 & -0.156 \\
        \hline
    \end{tabular}
\end{table}
\begin{figure}[H]
    \centering
    \includegraphics[width=0.5\textwidth]{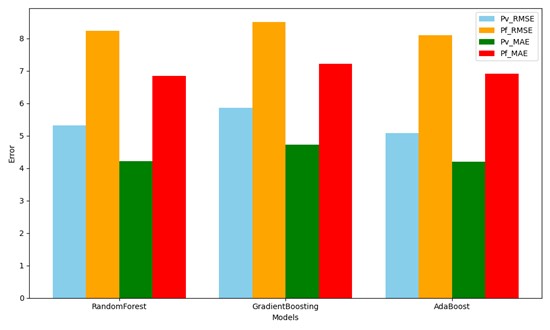}
    \caption{overall performance of Pv and Pf metrics across models.}
\end{figure}

\begin{table}[h]
    \centering
      \caption{Pv/Pf Prediction Summary (Random Forest)}
    \resizebox{\columnwidth}{!}{%
    \begin{tabular}{lcccccc}
        \hline
        Pf\_RMSE & Pf\_MAE & Pv\_MAE & Pv\_RMSE & Pv\_R\textsuperscript{2} & Pf\_R\textsuperscript{2} \\
        \hline
        5.311 & 8.232 & 4.220 & 6.834 & -0.188 & -0.270 \\
        \hline
    \end{tabular}%
    }
    \label{tab:pvpf_rf_summary}
\end{table}

\begin{table}[h]
    \centering
     \caption{Pv/Pf Prediction Summary (Gradient Boosting)}
    \resizebox{\columnwidth}{!}{%
    \begin{tabular}{lcccccc}
        \hline
        Pf\_RMSE & Pf\_MAE & Pv\_MAE & Pv\_RMSE & Pv\_R\textsuperscript{2} & Pf\_R\textsuperscript{2} \\
        \hline
        5.857 & 8.502 & 4.732 & 7.216 & -0.445 & -0.354 \\
        \hline
    \end{tabular}%
    }
    \label{tab:pvpf_gb_summary}
\end{table}

\begin{table}[h]
    \centering
        \caption{Pv/Pf Prediction Summary (AdaBoost)}
    \resizebox{\columnwidth}{!}{%
    \begin{tabular}{lcccccc}
        \hline
        Pf\_RMSE & Pf\_MAE & Pv\_MAE & Pv\_RMSE & Pv\_R\textsuperscript{2} & Pf\_R\textsuperscript{2} \\
        \hline
        5.072 & 8.095 & 4.204 & 6.915 & -0.084 & -0.228 \\
        \hline
    \end{tabular}%
    }
    \label{tab:pvpf_ab_summary}
\end{table}

\begin{figure}[H]
    \centering
    \includegraphics[width=0.5\textwidth]{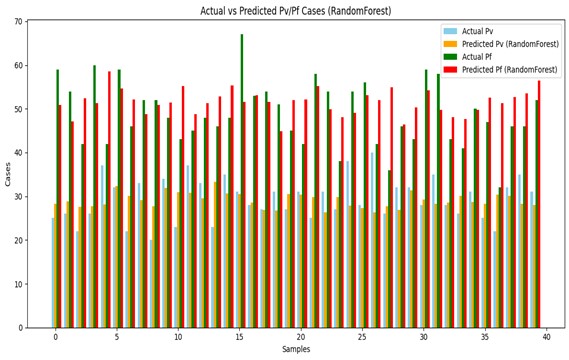}
    \caption{overall performance of Pv and Pf metrics across models.}
\end{figure}
\begin{figure}[H]
    \centering
    \includegraphics[width=0.5\textwidth]{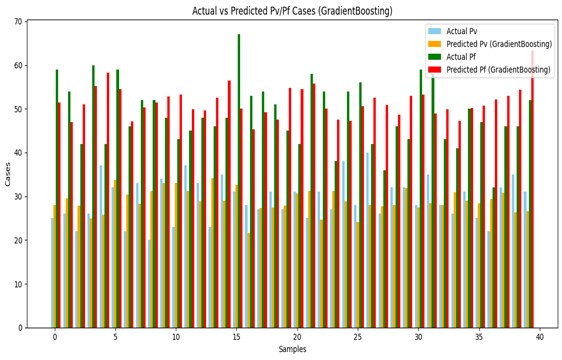}
    \caption{overall performance of Pv and Pf metrics across models.}
\end{figure}
\begin{figure}[H]
    \centering
    \includegraphics[width=0.5\textwidth]{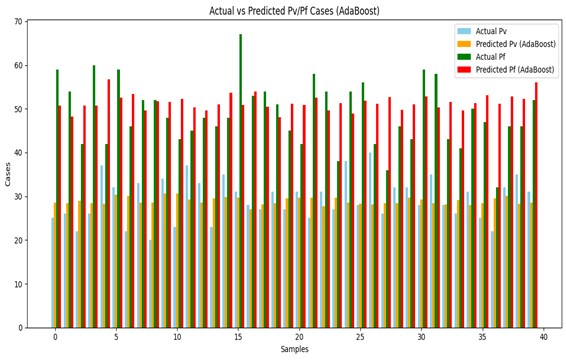}
    \caption{overall performance of Pv and Pf metrics across models.}
\end{figure}
\subsubsection{Discussion}
Malaria incidence can be predicted using the hybrid predictive modeling framework, which blends time-series forecasting, multi-output regression, and conventional regression-based prediction.
Compared to single-method approaches, the proposed model shows improved prediction accuracy, uncovers hidden patterns, and gives public health authorities valuable information.
In order to support evidence-based decision-making, targeted interventions, and resource optimization, the study provides a valid and repeatable predictive framework for malaria incidence in endemic areas.
This study proposes developing hybrid predictive models that integrate time-series forecasting and multi-output regression to bridge the gap in systematic modeling and prediction of malaria trends in the Amhara region.
By combining complementary strengths, the hybrid approach aims to improve predictive accuracy and generate insights that single-method models cannot provide.
Multi-output regression allows for the simultaneous prediction of species-specific case counts, spatial variations, and temporal dynamics.
Hospital Management Information Systems (HMIS) are present in the majority of healthcare facilities; however, their application in aiding decision-making and controlling malaria remains inadequate, particularly in the Amhara Region.
Previous research focuses primarily on identifying high-risk areas and does not provide useful insights into the disease's future occurrence.
Few all-encompassing models are able to forecast several outcomes simultaneously, such as cases specific to a given Plasmodium species, trends over time, and spatial variations.
By generating accurate predictions of future malaria trends, the model will provide health authorities with valuable information to guide prevention and control efforts.
To accurately predict the occurrence of malaria, the study will create a hybrid predictive model that combines multi-output regression and time-series forecasting.
By demonstrating the superiority of hybrid approaches and multi-output regression over traditional single method models, the study will advance methodological advancements in epidemiological predictive modeling.
With the aid of the predictive model's evidence-based insights, health officials in the Amhara Region will be able to prioritize interventions, allocate resources optimally, and reduce the burden of malaria.
Few studies have attempted to combine multiple modeling approaches, and hybrid predictive framework applications in malaria research are still rare.
To improve accuracy and provide valuable insights, predictive models that combine the benefits of multiple approaches are desperately needed.
The technique increases forecast accuracy and generalizability by combining multi-output regression, which enables the simultaneous prediction of multiple related targets, such as the prevalence of malaria in different districts or for different Plasmodium species.
Random Forest and AdaBoost, two ensemble learning techniques, are renowned for their robustness and ability to spot complex, nonlinear patterns in data. They can also increase model generalization and prediction accuracy.There are often no models that can simultaneously account for temporal and spatial dynamics, so the study aims to develop models that do so in order to better understand and predict patterns of malaria incidence across different regions and time periods. Using advanced modeling techniques that consider these dynamics aims to enhance forecasting capabilities and generate predictions that are more accurate and timely, which are essential for proactive efforts to control and eradicate malaria in the Amhara Region.
The predictive models are more applicable to regional health policy and intervention planning because the dataset captures the ecological and epidemiological characteristics of malaria transmission at the local level.
Because ensemble approaches reduce variance, enhance generalization, and accurately model nonlinear interactions between epidemiological, demographic, and environmental factors, they are particularly well-suited for epidemiological prediction tasks.
Using multi-output regression enables the model to account for the spatial-temporal dynamics of Plasmodium species-specific cases, resulting in a more thorough understanding of disease patterns in the context of malaria prediction.
\section{Conclusion}
To accurately predict the occurrence of malaria, the study created a hybrid predictive model that combines multi-output regression and time-series forecasting.
Unlike existing models, this framework allows for the simultaneous prediction of several outcomes, including Plasmodium species-specific cases, temporal trends, and spatial variations.
By demonstrating the superiority of hybrid approaches and multi-output regression over traditional single-method models, the study advances the methodology of epidemiological predictive modeling.
With the aid of the predictive model's evidence-based insights, health officials in the Amhara Region will be able to prioritize interventions, allocate resources optimally, and reduce the burden of malaria.
Because ensemble approaches reduce variance, enhance generalization, and accurately model nonlinear interactions between epidemiological, demographic, and environmental factors, they are particularly well-suited for epidemiological prediction tasks.
Using multi-output regression enables the model to account for the spatial-temporal dynamics of Plasmodium species-specific cases, resulting in a more thorough understanding of disease patterns in the context of malaria prediction.
The predictive models are more applicable to regional health policy and intervention planning because the dataset captures the ecological and epidemiological characteristics of malaria transmission at the local level.
\bibliographystyle{IEEEtran}
\bibliography{Reference.bib}
\end{document}